\documentclass{elsart}

\usepackage{amssymb}
\usepackage{graphicx}

\begin{document}

\begin{frontmatter}



\title{Reducing the magnetic susceptibility of parts in a magnetic
gradiometer}


\author[uwa]{Andrew Sunderland\corauthref{cor1}}\ead{asund@physics.uwa.edu.au}, \author[uwa]{Li Ju}, \author[uwa,gravitec]{Wayne McRae}, \author[uwa]{David G Blair}
\address[uwa]{School of Physics, University of Western Australia, Perth, WA, Australia}
\address[gravitec]{Gravitec Instruments, Perth, WA, Australia}
\corauth[cor1]{Corresponding author. Tel.: +61 422 282 438.}

\begin{abstract}
% Text of abstract
In this paper we report a detailed investigation of a number of
different materials commonly used in precision instrumentation in
the view of using them as critical components in the magnetic
gradiometer. The materials requirement inside a magnetic gradiometer
is stringent because the presence of magnetic susceptible material
will introduce intrinsic errors into the device. Many commercial
grade non-magnetic materials still have unacceptably high levels of
volume magnetic susceptibility between $10^{-3}$ and $10^{-4}$. It
is shown here that machining with steel tools can further increase
the susceptibility by up to an order of magnitude. The ability of an
acid wash to remove this contamination is also reported. Washing in
acid is shown to reduce the variation of volume susceptibility in
several commercial grade plastics which already have low values of
susceptibility.
\end{abstract}

\end{frontmatter}

% main text
\section{Introduction}

\subsection{Magnetic Gradiometers}

A major application for magnetic gradiometers is measuring the
gradient of the magnetic field produced by nearby geological targets
at distances of 30 m to 100 m\cite{mushayandebvu}. Magnetic material
may also be present inside the gradiometer, inside attached
equipment and cables, or inside the vehicle used to deploy the
gradiometer. All of these pieces of equipment will produce their own
magnetic gradients. Whereas magnetic fields from far field dipole
sources scale as the inverse third power of distance, magnetic
gradients scale as the inverse fourth power and are particularly
sensitive to close objects.

Recently, a Direct String Magnetic Gradiometer (DSMG) has been
developed, which employs a single "string" as the sensing
element\cite{golden}. The use of a long and thin sensing element in
the DSMG means that a significant number of instrument components
are near the sensor string where even relatively small levels of
magnetic contamination can lead to false signatures many times
larger than the signal from the distant geological target.

The magnetic material inside the sensor generates both induced and
remanent magnetisation. The later produces a constant magnetic
gradient that can be subtracted from the signal without impeding the
operation of the magnetic gradiometer whereas induced magnetisation
aligns in the Earth's magnetic field. If the sensor is rotated while
being deployed in a moving vehicle, the magnetic gradient produced
by induced magnetisation will vary. This varying distortion in the
measured magnetic gradient is called heading error. Here we report
an investigation of reducing susceptibilities of different materials
with a view to reducing the amount of heading error in this
particular type of magnetic gradiometers.

\subsection{Reasons for measuring the magnetic susceptibility}

For small values of magnetic field, the magnetisation rises linearly
with applied magnetic field strength. This means that the induced
magnetisation per unit volume from a component is equal to the
volume susceptibility of the material multiplied by the strength of
the applied field (which in this case is the Earth's 30 000 nT to 60
000 nT field). The magnetic field in the space surrounding a
magnetised object scales as the inverse third power of distance.
Therefore induced magnetisation will produce a heading error that
will depend on the distance of the component from the sensing
element of the magnetic gradiometer, the volume of the component,
any anisotropy in the shape of the component, and the volume
magnetic susceptibility of the component.

There is more data in the literature on the remanent magnetic moment
of acid washed samples (see below). Nevertheless, in this paper
volume susceptibility is required to quantify the amount of heading
error a material will produce.

\subsection{Prior work}

Measurements of the magnetic susceptibility before and after an acid
wash have been well known in the literature, see for example Spencer
and John\cite{spencer}. Spencer and John washed their samples in
hydrochloric acid to remove possible traces of iron and repeated
until a constant value was obtained for the susceptibility. The acid
wash was incidental to Spencer's paper and the actual susceptibility
values were not reported. Much advice about avoiding magnetic
effects in industry is based on oral tradition (for example use
phosphor bronze and there will be no problem).

Honda measured the mass magnetic susceptibility of various pure
metals, in different forms such as ingot, wire or cast\cite{honda}.
Honda also recorded the concentration of iron and the chemical form
that the iron impurity takes inside the base metal. Honda found only
insignificant variation in susceptibility between different metal
forms.

Constant and Formwalt measured the remanent magnetic moment of a
series of metals\cite{constant}. Measurements were taken of both the
commercial grade metal and the chemically pure metal. Commercial
brass was reported to have the highest magnetic moment, followed by
commercial copper and silver. More recently Keyser and
Jefferts\cite{keyser} measured the magnetic susceptibility of a wide
variety of laboratory construction materials.

Measuring the volume magnetic susceptibility is not the optimum way
to identify small amounts of ferromagnetism because the magnetic
susceptibility in the material could be due to diamagnetism,
paramagnetism as well as ferromagnetism. On the other hand,
measuring a non-zero remanent magnetic moment when the external
field is zero is a definite indication of ferromagnetism (although
soft iron ferromagnets can have high susceptibility and near zero
remanence). Remanent magnetic moment measurements are the principal
method of checking a sample for magnetic contamination, see for
example Matsubayashi et al\cite{matsubayashi} or Wang et
al\cite{wang}.

\section{Typical magnetic materials}

The three ferromagnetic elements (Fe, Ni and Co) have very high
susceptibilities, the highest being the initial relative magnetic
permeability of 99.9\% pure iron $\mu_{r} = 25000$\cite{mccurrie}.
When trying to reduce the magnetic contamination it is not enough to
merely avoid the use of pure iron, nickel or cobalt parts because
other nonferrous metals of commercial grade are often less than 99\%
pure and usually contain iron as an impurity.

Brass is often used as a nonmagnetic substitute to replace iron in
parts that require high strength or density\cite{brassapplications},
but the magnetic volume susceptibility of brass varies considerably
and care must be taken when choosing a supplier. Generally volume
susceptibility will increase with increasing iron content, although
the susceptibility will be higher if the iron impurity is
concentrated in small clumps\cite{huck} or if precipitation of iron
occurs during heat treatment\cite{butts}. Small amounts of iron
(\texttt{<} 0.05\%) can alloy with the copper in brass to produce an
alloy with a volume magnetic susceptibility proportional to the
square of the iron concentration\cite{ekstrom}.

\begin{figure}[ht] %h here, t top, b bottom,  p page, ! as soon as possible
\centering
\includegraphics[width = 0.5\textwidth]{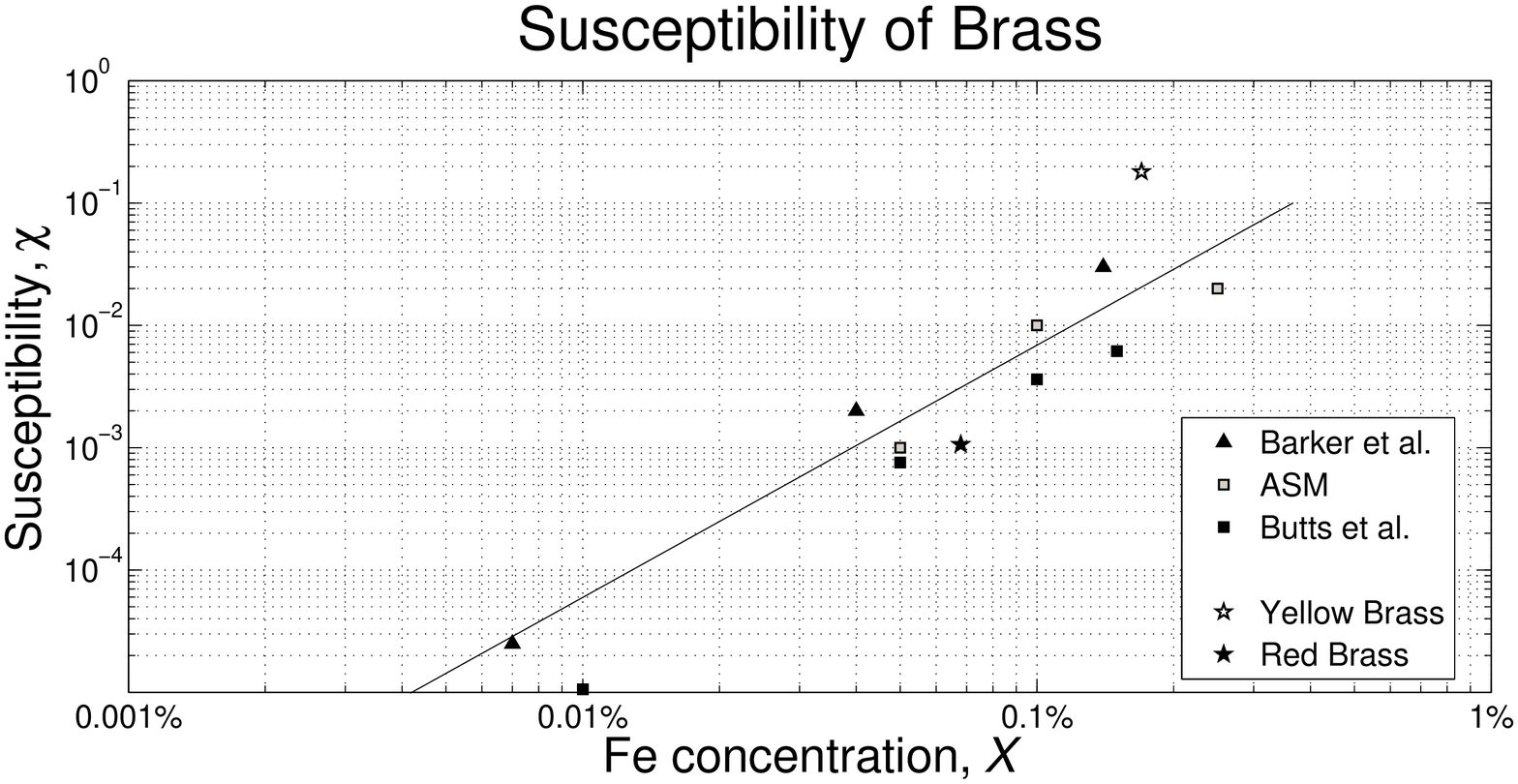}
\caption{The susceptibility of yellow brass rises rapidly with
increasing iron concentration. The data is compiled from a paper by
Barker et al\cite{barker}, a book by ASM International\cite{asm}, a
paper by Butts et al\cite{butts} and measurements performed by the
authors on yellow and red brass samples. The line is a best fit at
low Fe concentrations using a square law $\chi \propto
\textrm{X}^{2}$.} \label{brass}
\end{figure}

The complete analysis of the susceptibility of brass is quite
complicated as other factors such as heat treatment, cold working
and oxygen concentration can have a large effect, see for example
Fickett and Sullivan\cite{fickett}. Parts inside the DSMG are
expected to be exposed to large temperature extremes during
construction and operation. Relying on a heat treatment to lower the
susceptibility of a magnetic gradiometer is not sufficiently robust
for all environments. For this reason the susceptibility values of
prior work quoted in this paper are all referring to the
susceptibility of the material as cast or formed which tend to be
higher than textbook values. Fig. \ref{brass} shows some previous
work on volume susceptibility vs. iron concentration for yellow
brass (60-70\% Cu, 30-40\% Zn). The graph shows a good fit with the
square law at low iron concentrations.

If an isotropic very low magnetic susceptibility is required then
metals with more than 0.01\% iron should not be used. Unreinforced
plastics are the best materials to use in extreme nonmagnetic
conditions since they have very low levels of
impurities\cite{goodfellow}. The three unreinforced plastics (Torlon
4203, PET and PTFE) in Table \ref{plastictable} all have less than
0.0001\% iron. The composite materials (Torlon 4301 and G10) and the
ceramic (macor) have higher impurity levels although not as high as
the impurity levels of the metals. The exception is 99.95\% pure
oxygen free highly conductive copper. Pure metals can have lower
impurity levels than commercially available alloys\cite{goodfellow}.

\section{Hacksaw contamination}

In addition to the volume susceptibility of the bulk of the
material, there can also exist residual contamination from the
machining of parts that produces a significant surface contribution
to induced magnetisation. To investigate this surface contribution,
14 samples of 7 different materials were cut to dimensions 12 mm x
16 mm x 16 mm with a high carbon steel hacksaw. In addition two M3
holes were tapped with a tap made from tool steel. Depending on the
abrasiveness of the sample in question, some of the steel on the
tools will be deposited into the surface of the samples during
machining. Table \ref{toolstable} shows that the initial relative
magnetic permeability of a hacksaw blade can be as high as $\mu_{r}
= 11$ compared to magnetic susceptibilities of $\chi_{diamagnet}
\approx -10^{-5}$ for a typical diamagnet. This means that even
small amounts of steel contamination can produce unacceptably high
heading error.

To remove any surface magnetism, the samples were immersed in 3\%
concentrated hydrochloric acid. Low field volume susceptibility
readings were taken before and after immersion using a Bartington
MS2b susceptibility meter for the metal samples and a ZH Instruments
SM-30 susceptibility meter for the plastic and ceramic samples. The
DSMG operates at room temperature, hence all susceptibility
measurements in this paper were taken at room temperature. All
susceptibility values are volume susceptibility in SI (MKS) units.
Concentrations in samples are stated by mass.

\begin{figure}[ht] %h here, t top, b bottom,  p page, ! as soon as possible
\centering
\includegraphics[width = 0.5\textwidth]{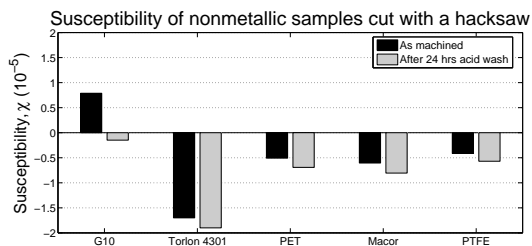}
\caption{Low field ( \texttt{<} 50 $\mu$T) susceptibility of plastic
and ceramic samples before and after a acid wash. Data taken using a
ZH Instruments SM-30 susceptibility meter which has uncertainty of
$\pm 10^{-6}$ for samples of this size. An adjustment for sample
volume in the SM-30 was done using a formula derived by Gattacceca
et al\cite{gattacceca} which has an uncertainty of $\pm
20\%$.}\label{plasticsusceptibility}
\end{figure}

Fig. \ref{plasticsusceptibility} shows the change in volume
susceptibility before and after acid treatment for each of the
plastic and ceramic samples. The graph shows a slight decrease in
the volume susceptibility of the plastic and ceramic samples after a
24 hour acid wash. There is very little contamination to remove from
any of the plastic and ceramic samples, with the exception of G10
(G10 is significantly more abrasive than the other plastics and
removes more iron from tools during machining). The advantage of
immersing the plastic samples in acid was that the volume magnetic
susceptibility had less variation after the acid wash and that the
magnetisation was more isotropic. The volume susceptibility of
samples as machined varied $\pm 2 \times 10^{-6}$ when rotated to
different orientations whereas the volume susceptibility of acid
washed samples varied only $\pm 10^{-6}$. Hydrochloric acid removes
surface iron from all samples, but when it is applied to metallic
samples hydrochloric acid may also corrode the surface of the metal.
To evaluate the rate that hydrochloric acid corroded the metal
samples, each of the samples was immersed in hydrochloric acid for
durations of 10 minutes, 70 minutes, and 24 hours. Table \ref{mass}
shows some loss of mass from acid washing. In particular the
corrosion in the aluminum sample after 24 hours produced a
significant 7\% reduction in the mass.

\begin{figure}[ht]%h here, t top, b bottom,  p page, ! as soon as possible
\centering
\includegraphics[width = 0.5\textwidth]{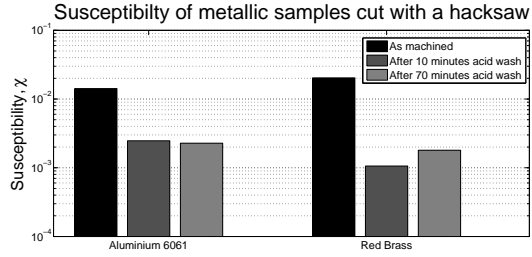}
\caption{Low field ( \texttt{<} 250 $\mu$T) susceptibility of
metallic samples before and after acid washes. Data taken using a
Bartington MS2b susceptibility meter which has uncertainty of $\pm
10^{-5}$ for samples of this size.}\label{metalsusceptibility}
\end{figure}

Fig. \ref{metalsusceptibility} shows that immersing the metallic
samples in acid for only 10 minutes removes nearly all the surface
contamination and reduces the volume susceptibility by an order of
magnitude. An acid wash of 70 minutes produces no further reduction
in the volume susceptibility. From this result it can be inferred
that the remaining induced magnetism is caused by small but
significant concentrations of iron in the body of the sample. A 10
minute acid wash is therefore optimal for reducing heading error
from brass and aluminium parts because immersing parts in acid for
longer durations of time may corrode parts to the extent that they
cease to be within design tolerances.

The volume susceptibility of the metallic samples before acid
washing are approximately 3 orders of magnitude higher than the
plastic samples before acid washing. This indicates that the more
abrasive metal samples are more easily contaminated with the high
carbon steel in the hacksaw.

The value of the volume magnetic susceptibility of the acid washed
aluminium 6061 sample measured by the authors does not agree with
previous work by Keyser and Jefferts\cite{keyser}. This could be due
to different amounts of magnetic contamination between samples or an
erroneous response of the Bartington MS2 to electrical conductivity
in the sample as shown in Benech and Marmet\cite{benech}. However,
for the purpose of this paper, in order to avoid any possible
ambiguity in getting samples with different contamination levels,
the use of such materials should be avoided. The red brass
susceptibility value recorded by the authors does agree with
previous work as shown in Fig. \ref{brass}.

\section{Lathe contamination}

10 samples of 5 different materials, were cut on a lathe with a tool
bit made from high speed steel. The samples were machined on the
lathe to make cylinders of diameter 5 mm and length 5 mm in order to
fit in a Vibrating Sample Magnetometer (VSM). A single M3 hole was
tapped along the axis with a tool steel tap. The 8 metal samples
were immersed in acid for 10 minutes and the 2 Torlon samples were
immersed for 24 hours. Volume susceptibility measurements were taken
before and after immersion following the same procedure used for the
larger samples.

\begin{figure}[ht]%h here, t top, b bottom,  p page, ! as soon as possible
\centering
\includegraphics[width = 0.5\textwidth]{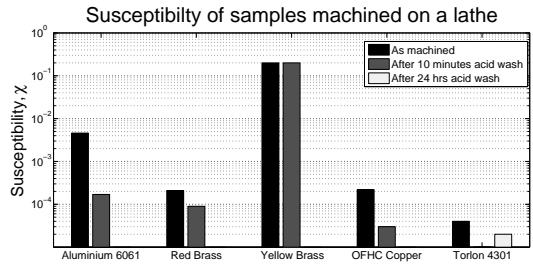}
\caption{Low field ( \texttt{<} 0.02 T) susceptibility of
cylindrical samples before and after acid washes. Data taken using a
Aerosonic 3001 Vibrating Sample Magnetometer which has uncertainty
of $\pm 10^{-5}$ for samples of this
size.}\label{cylindersusceptibility}
\end{figure}

Fig. \ref{cylindersusceptibility} shows the change in volume
susceptibility before and after acid treatment for each of the
samples. The smaller cylindrical samples should have a higher
relative surface contamination due to the large surface to volume
ratio. However, the volume susceptibility of the unwashed aluminium
cylindrical sample is a factor of three less than its hacksawed
counterpart. This is most likely because the relative permeability
of the lathe tool bit is only $1.4$ compared to a relative
permeability of $11$ for the hacksaw blade. Another factor is that
the lower hardness of the carbon steel hacksaw blade listed in Table
\ref{toolstable} will allow the blade to deposit more steel onto the
surface of the sample

The unwashed red brass and copper cylindrical samples have almost
the same value for volume susceptibility of $2.2 \times 10^{-4}$.
This value is two orders of magnitude lower than the hacksaw
contaminated brass sample. This indicates that copper and copper
alloys are difficult to contaminate with high speed steel.

The yellow brass sample has a very high bulk susceptibility and the
percentage change after an acid wash was unmeasurably small. The
volume susceptibility of the washed cylindrical samples was lowest
for the materials with the least amount of iron. The OFHC copper
sample with 0.0002\% Fe has a volume susceptibility of $3 \times
10^{-5}$ and Torlon 4301 with 0.0002\% Fe has a volume
susceptibility of $2 \times 10^{-5}$.

There is a discrepancy between the volume magnetic susceptibility of
OFHC copper measured using the VSM and the textbook value of $-1
\times 10^{-5}$\cite{asm}. This could be due to an imperfect acid
wash or the limited accuracy of the VSM of $\pm 10^{-5}$ for samples
of this size. Despite this inaccuracy, the results of the VSM are
show that an acid wash can reduce the volume suspectibility below
$10^{-4}$ which is sufficient for reducing heading error.

\section{Varying applied field measurements}

In addition to the low field volume susceptibility measurements, a
complete scan of the magnetisation at different applied fields up to
$1$ T was preformed on the two 5 mm diameter Torlon cylinders. Two
additional measurements were made at $-7$ T and $7$ T to check the
diamagnetic contribution. The measurements were taken using a
Quantum Design MPSM-7 Superconducting Quantum Interference Device
(SQUID) before and after immersion in acid.

\begin{figure}[ht]%h here, t top, b bottom,  p page, ! as soon as possible
\centering
\includegraphics[width = 0.5\textwidth]{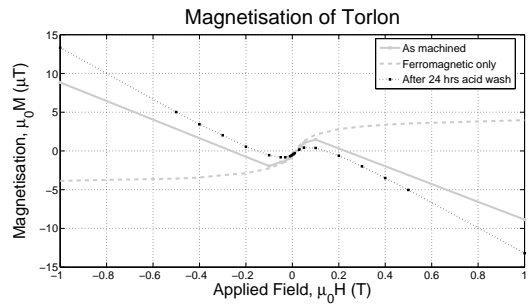}
\caption{Magnetisation of Torlon sample 2 in applied fields from -1
T to 1 T before and after acid washes. The dashed line is an
inferred curve of the ferromagnetic contribution to the
magnetisation of the acid washed samples. Data taken using a Quantum
Design MPSM-7 SQUID. The SQUID has a precision of $\pm 0.2$ nT
magnetisation for samples of this size although the accuracy is $\pm
5\%$ due to imperfect alignment of the
samples.}\label{torlonsusceptibility}
\end{figure}

From the gradient at the origin (low field case) in Fig.
\ref{torlonsusceptibility}, the volume susceptibility of the
unwashed and washed samples are both positive with values of $3.6
\times 10^{-5}$ and $2.0 \times 10^{-5}$ respectively. When the
applied magnetic field exceeds $\mu_{0}H > 0.1$ T the magnetisation
has saturated and the volume susceptibility becomes negative. The
high applied field susceptibilities for the unwashed and washed
samples are $-1.4 \times 10^{-5}$ and $-1.7 \times 10^{-5}$
respectively. These values indicate that Torlon is diamagnetic so
that a pure sample with no iron contamination should have a volume
susceptibility near $-1.7 \times 10^{-5}$ in both low and high
applied fields. The presence of a positive susceptibility that
saturates in a modest applied magnetic field indicates the presence
of a small amount of ferromagnetism.

Immersing the sample in acid reduced the low field volume
susceptibility from $3.6 \times 10^{-5}$ down to $2.0 \times
10^{-5}$. The acid wash removed all surface contamination and
reduced the ferromagnetic contribution of the entire sample by 30\%.
The remaining ferromagnetism in the washed sample is produced by the
bulk of the sample.

After subtracting the diamagnetic contribution to magnetisation,
what is left is the magnetisation from ferromagnetic sources alone.
Fig. \ref{torlonsusceptibility} shows that this ferromagnetic
contribution in the washed Torlon saturates at a magnetisation of
$\mu_{0}M = 4.1$ $\mu$T. By comparison pure iron saturates at a
magnetisation of $\mu_{0}M = 2.15$ T, nickel saturates at $\mu_{0}M
= 0.69T$ and cobalt saturates at $\mu_{0}M = 1.79$ T\cite{zhou}.
Assuming the ferromagnetic contribution is all due to Fe, the
ferromagnetic iron in this Torlon 4301 sample is about $10$ parts
per million by mass. The result of 0.086\% Fe from elemental
analysis suggests that almost all of the iron is in a paramagnetic
state, a consequence of the iron being evenly dissolved into the
Torlon.

There is a discrepancy between the low field susceptibilities of the
washed 5 mm diameter Torlon cylinder and the susceptibility of the
washed larger 12 mm x 16 mm x 16 mm prism. Possible causes could be
a large anisotropy in the induced magnetism or that the acid is not
removing all of the surface ferromagnetism which is more significant
in the smaller samples. Despite the discrepancy, the change in
volume susceptibility after an acid washed was negligible for both
the cylinder and the prism.

\section{Conclusion}
With regard to minimizing heading error in a magnetic gradiometer,
plastics and ceramics parts that were measured had the lowest values
for volume susceptibility $\chi_{plastic} \sim 10^{-5}$. In addition
the effect of machining plastic with steel tools produced negligible
contamination. Metallic parts may however be required for their high
conductivity or strength. Metals with less than 0.01\% iron, which
have been acid washed can also have susceptibilities below
$10^{-4}$, comparable with plastics. Machining metals with high
speed steel is preferred to high carbon steel as it produces less
contamination. Metallic samples should be immersed in 3\%
concentrated hydrochloric acid for an optimum time of 10 minutes.

Washing plastic or ceramic parts in acid will reduce the volume
susceptibility by only a insignificant amount. Out of the acid
washed samples, the volume magnetic susceptibility closest to zero
was $-6 \times 10^{-6}$ from PTFE. Susceptibilities significantly
lower than that of PTFE in solid parts can only be achieved by using
specially designed alloys that offset paramagnetism with
diamagnetism. Nevertheless using expensive alloys is not necessary
since even assuming maximum asymmetry, volume susceptibilities of
order $10^{-5}$ will produce heading error in the DSMG of order $10$
nT/m peak which is less than the typical heading error of $20$ nT/m
peak from the vehicle used to deploy the gradiometer. The anisotropy
in the volume susceptibility of acid washed plastics is only $\pm
10^{-6}$. A reduction in the anisotropy of magnetic impurities
coupled with symmetry in the sensor could reduce heading error. We
intend to investigate this in future work.

\ack The authors would like to thank Dr. Alexey Veryaskin and Mr.
Howard Golden of Gravitec Instruments for many useful discussions
and suggestions, A/Prof. Tim St Pierre of the BioMagnetics and Iron
BioMineralisation group at UWA for the use of the wet acid
laboratory, Dr. Robert Woodward of the Nanomagnetics and Spin
Dynamics Group at UWA for the SQUID and VSM measurements, Prof. Li
of the Tectonic Special Research Center at UWA for the use of a MS2b
susceptibility meter, Mr. Barry Price of the Chemistry Centre of WA
for elemental analysis and Mr. Mads Toft of Alpha Geoscience for the
use of a SM-30 susceptibility meter. Work on the DSMG project is
funded in part by a linkage grant from the Australian Research
Council.

% \appendix

% \section{}
% \label{}

\begin{table}[ht]
\centering
\includegraphics[width = 0.46\textwidth]{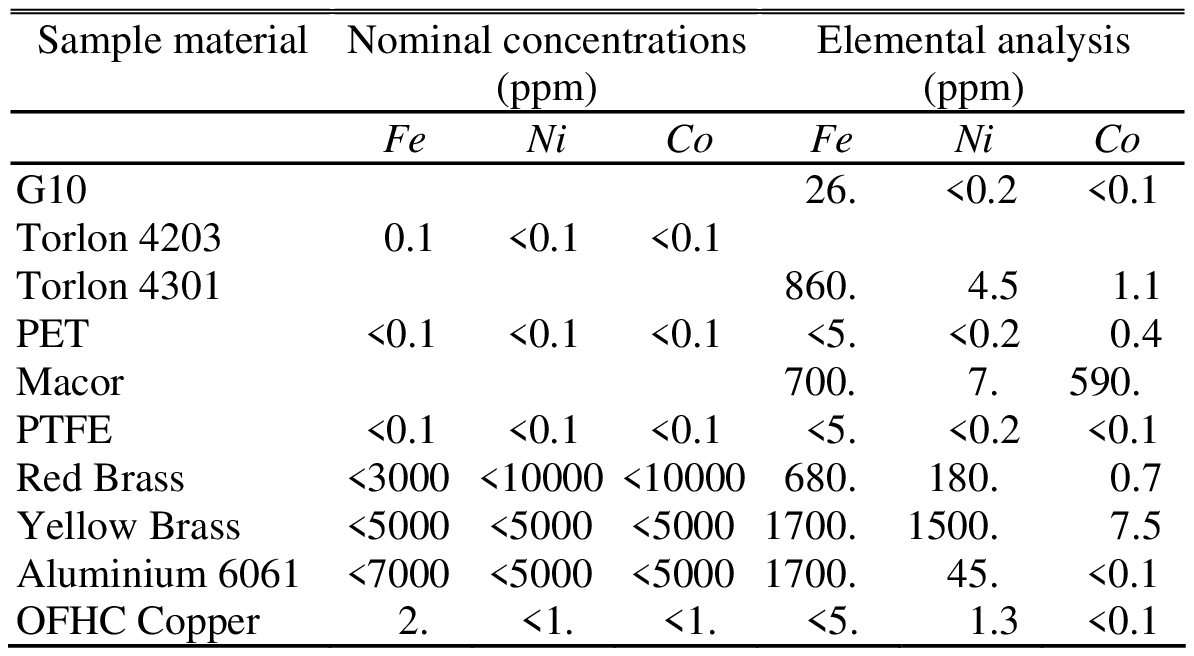}
\caption{\rm{Nominal concentrations were taken from material
datasheets. Elemental analysis was performed by the Chemistry Centre
of Western Australia using a Inductively Coupled Plasma - Atomic
Emission Spectrometer (ICP-AES).}}\label{plastictable}
\end{table}

\begin{table}[ht]
\centering
\includegraphics[width = 0.5\textwidth]{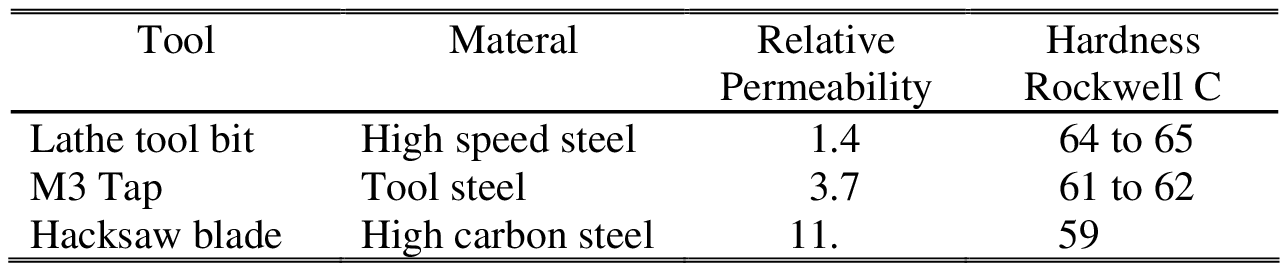}
\caption{\rm{The permeability measurements were made with a
Bartington MS2b susceptibility meter. The permeability of the tools
varied $\pm 30\%$ when rotated. Hardness values are from a book by
ASM International\cite{asm2}.}}\label{toolstable}
\end{table}

\begin{table}[ht]
\centering
\includegraphics[width = 0.5\textwidth]{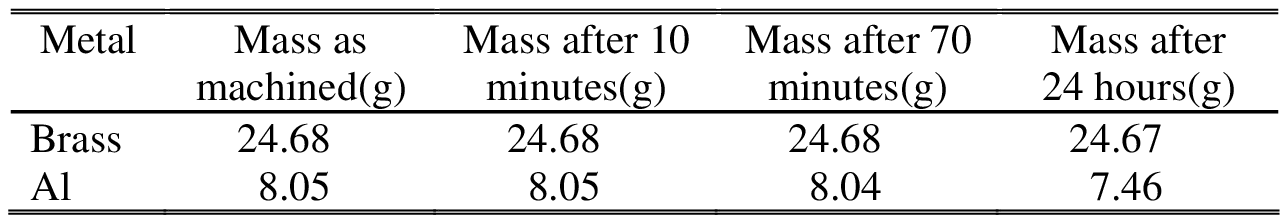}
\caption{\rm{The mass reduction of metal samples in acid increases
with longer immersion times, uncertainty is $\pm$ 0.01g.}}
\label{mass}
\end{table}

\end{document}